\documentclass[preprint,aps,epsfig]{revtex4}
\usepackage{graphics}
\usepackage{graphicx}

\begin{document}
\begin{center}
{\Large \bf Cross-over behaviour in a  communication network}
\end{center}
\begin{center}
Brajendra K. Singh$^1$\hspace{0.1in} and \hspace{0.1in} Neelima Gupte$^2$\\
Department of Physics, Indian Institute of Technology, Madras,\\
Chennai, 600 036
\end{center}

\begin{center}
{\bf \large Abstract}
\end{center}
{
We address the problem of message transfer in a communication network. The
network consists of nodes and links, with the nodes lying on a two dimensional
lattice. Each node has connections with its nearest neighbours, whereas some
special nodes, which are designated as hubs, have connections to all the sites
within a certain area of influence. The degree distribution for this network is
bimodal in nature and has finite variance. The distribution of travel times
between two sites situated at a fixed distance on this lattice shows
fat fractal behaviour as a function of hub-density. If extra assortative
connections are now introduced between the hubs so that each hub is connected
to two or three other hubs, the distribution crosses over to power-law 
behaviour. Cross-over behaviour is also seen if end-to-end short cuts are
introduced between hubs whose areas of influence overlap, but this is much
milder in nature. In yet another information transmission process,
namely, the spread of infection on the network with assortative connections,
we again observed cross-over behaviour  of another type, viz. from one
power-law to another for the threshold values of disease transmission
probability.  Our results are relevant  for the understanding of the role of
network topology in information spread processes.
}
\noindent e-mail:$^1$braj@chaos.iitm.ernet.in, $^2$gupte@chaos.iitm.ernet.in  
\newpage
\section{Introduction}
In recent years there has been an unprecedented rise of interest in the study
of networks and their properties \cite{Newman, Barabasi}. These networks can be
basically regarded as a  collection of nodes linked by  edges. The nodes
represent individual entities, while the edges  represent  interaction
between  any pair of nodes in the network. Many natural systems, as well as
engineered systems, can be represented by such networks. Some examples of such
networks  are found at large spatial scales, $e.g$, the internet or the
power grid, while others are at single organismic levels, $e.g$, metabolic
pathways \cite{Newman, Barabasi, Amaral}. The structure and connectivity
properties of such networks and the capacity and degree of connectivity of
their nodes have important consequences for processes which occur on the
network. Some examples of such processes are message transfer in communication
networks, the spread of infectious diseases in social networks, the spread of
computer viruses in computer networks and avalanche-spread in load-bearing
networks \cite{Watts-newman,Moore,Kuperman,Ziff,Newman1,Janaki}.
Such network studies have been seen to have potential applications in different
fields of science and technology, including epidemiology and ecology.\\

Several different classes of network topologies have attracted recent interest.
These include small{-}world (SW) network models which have short paths
between any two nodes and highly clustered connections \cite{Strogatz},
and their variants \cite{Watts-newman,Ziff}, generalised families of
small-world networks \cite{Kleinberg} and scale{-}free (SF) networks 
which incorporate the existence of `hubs' viz. nodes  with many more
connections than the average node\cite{ALB-Reka,ALB-Reka-Jeong}. There have
also been attempts to characterise the topology of natural networks, using
measures like average path lengths, clustering co-efficients, degree
distributions and the existence of `network motifs'\cite{Milo}, viz. 
patterns of inter-connections occurring in complex networks at numbers that are
significantly higher than those in randomised networks.\\

It is known that the nature of the network topology significantly affects the
manner in which spread processes occur on networks. A striking example of this
was found by Kleinberg, where a decentralised algorithm was able to find very
short paths (resulting in very short delivery times for messages) for a two
dimensional network where long range connections followed the inverse square
law. Another example where the topology of the network had a significant role
to play is in the spread of computer viruses. It was seen that computer viruses
can spread on the SF networks with zero transmission threshold
\cite{Vespignani1,Vespignani2} and that networks of differing topologies have
differing degrees of vulnerability to attack \cite{Reka-Jeong-Barabasi,DSCall,
Cohen-Erez} and different extents of error tolerance. The importance of such
studies from the point of view of applications, is obvious.\\

In this paper, we set up a communication network on a two dimensional lattice.
Communication networks base on two dimensional lattices have been considered
earlier in the context of search alogorithms \cite{Kleinberg}, as well as in
the context of traffic on lattices with hosts and routers  \cite{Ohira,Sole,
Fuks1, Fuks2, FN}. The lattice we consider in this paper has two types of
nodes, viz. regular nodes and hubs. Each regular node on the lattice has
connections with its nearest neighbours, whereas some special nodes, which 
are designated as hubs, have connections to all the sites within a certain
area of influence. Thus, nodes falling within the influence area of a hub can
directly interact with the hub, or {\it vice-versa}. The degree distribution
of this network is bimodal and has finite variance. We examine two distinct
spread  processes on this lattice network as functions of hub-density. 
The distribution of travel times for a message transmitted between a fixed pair
of sites for this lattice shows fat fractal behaviour as a function of
hub-density. If extra connections are now introduced between the hubs so that
each hub is connected to two or three other hubs, the distribution crosses over
to power-law behaviour. The end-to-end short cutting of hubs whose influence
areas overlap also induces cross-over, but this is much milder in nature.
We also examined the process of disease transfer on this network, a 
process that differs from message transfer in the fact that it is
an undirected probabilistic process and not a directed transfer between a
source and a target. Here again the transmission threshold plotted 
as a function of hub-density crossed over from one type of  power-law behaviour to another on the introduction of two extra connections per hub.

In  section ~\ref{Model} we set up the model of the communication network.
In the next  section (Section ~\ref{MT}) we discuss the problem of directed
message transfer between source and target nodes on this network.
In section ~\ref{Disease} we discuss  the spread of information  as a random
process as in the spread of disease.  The last section summarises our
conclusions.\\ 

\section{The model}
\label{Model}
As stated in the introduction, the proposed network is a two{-}dimensional
lattice of nodes where every node is connected to its nearest-neighbours. 
A certain fraction of the total nodes  are designated as hubs, viz. as nodes
which have connections to all nodes within their area of influence, where
the area of influence is defined as a square area around the hub, which
accommodates $(2k+1)^2$ nodes, including the hub node,  $k$ being the distance
in lattice units (LU) from the hub to the outer most node in any principal
direction within the square. All nodes within the influence area of a hub are
termed its constituent nodes. A schematic representation of a hub node is
shown in Fig.~1. These hubs are randomly chosen on the lattice maintaining
a minimum distance, $d_{min}$, between any pair of hubs. This parameter
($d_{min}$) controls the extent of overlap between the influence regions of
the hubs. See the two hubs, `d' and `e', in Fig.~1. The distance between them
is unity, and their influence regions are maximally overlapped. 
As the number of hubs on the lattice goes up, the number of connections that a
constituent node can have also goes up, as it increasingly acquires links with
more hubs because of the overlapping of the influence regions of the hubs.
Thus the model takes into account the fact that local clustering
in a geographical neighbourhood can occur for many realistic models.\\

The hubs in this model connect to a fixed number of nodes, as in many realistic
environments, the number of links that can attach to a given hub is limited by
the number of ports to the hub. We examine the influence that the presence
of hubs exerts on the connectivity distribution of the network. In the absence
of the hubs each node has the same degree of connectivity and
the degree distribution (where the degree of a node is the number of
nodes it is connected to) is a delta function with a single peak at four.
However, the introduction of hubs leads to completely different connectivity
patterns for the network. Due to the presence of  hubs, the connectivity of
the nodes ranges from four (for the regular nodes) to $(2k+1)^2-1$ (for the
hubs). The degree of the constituent nodes lies between the two extremes,
being equal to $C+4$ where the constituent node  lies in the overlapping
influence area of $C$ hubs. The degree of the constituent nodes is thus a
function of the hub density, which controls the extent to which the influence
areas of different hubs overlap. Fig.~2 shows a typical distribution of
the degree of connectivity of nodes of all types for a  lattice of size
$100\times100$, $k=3$, $d_{min}=1$ and the hub density is $4.0\%$.
As expected, there are two peaks in the distribution, one, a sharp one,
at $(2k+1)^2-1$ corresponding to the degree of connectivity of the hubs,
and the other around four (the degree of the regular sites) with a spread
around the value. This spread comes from the constituent nodes of hubs with
overlapping regions of influence. It is clear that  the degree distribution
does not fall into either of the two broad classes of networks -- the SW and
SF networks. It also differs from the connectivity patterns of the random
graphs. (See ref \cite{ALB-Reka-Jeong} for the connectivity distributions
for the SW and  SF networks as well as those of random graphs, and
\cite{warren,rozenfeld} for scale-free models with local clusters and
geographic separation.) The variance in the degree distribution of
our model is finite unlike that in the case of scale-free networks
where the variance is infinite. This fact can have very important
consequences for information spread processes on a network. The importance
of the variance in the spread of disease has already been noted in the case
of scale-free networks. It is therefore interesting
to investigate spreading processes in  models like ours which
do not have infinite variance, and to compare them with spreading processes
seen in the case of other networks, e.g. scale free networks with and without
local clustering in a geographic neighbourhood.\\

We now study  two types of communication problems in this network viz.  the
problem of message transfer and that of the spread of infectious disease.

\section{The problem of message transfer}
\label{MT}
We study the transfer of a message from an arbitrary source node to an
arbitrary target node on the lattice. Each (ordinary) node transfers the
message to the node nearest to it in the direction which will minimise the
distance from the current message holder to the target. When any constituent
node of a hub is the current message holder, then the node directly sends
it to the hub. The lattice distance between the sender node and the hub is
just one hop because of direct communication between them, thereby speeding
up the process of message transmission. If the hub is the current message
holder the message is forwarded to one of the constituent nodes within its
influence area, the choice of constituent node being made by minimising the
distance to the target. Thus the presence of hubs on the lattice increases the
message transfer speed along the path, and the total travel time depends
primarily on how many hubs  fall on the path for the given influence radius,
$k$. A typical travel path for such a lattice is shown in Fig.~1 and is also
indicated by the label  $`O'$ in Fig.~3.

\subsection{Speed enhancement schemes}
It is clear that the presence of hubs on the lattice makes the process of the
message transfer faster than the situation when there are no hubs. However,
there is further scope to enhance the speed of transmission. Here we discuss
two speed enhancing schemes which can be practicably applicable to the
communication process.\\
\noindent
{\bf Scheme~I}: 
Whenever there is  an overlap between influence areas of hubs, then the
message is transferred from the first hub to second then to the third, so on.
Intermediate constituent nodes do not participate in the process of message
transfer towards the target. For example, if there is an overlap between the
influence regions of two hubs, namely, {\it A} and {\it B}, along the path,
then after receiving the message the first recipient constituent node of
{\it A} directly forwards it to its hub, {\it A}, which directly sends it to
{\it B}, from which the message is subsequently sent to the second recipient
constituent node of {\it B}. The second recipient node of {\it A} and the
first recipient node of {\it B} here do not take part in the message transfer. 
This scheme, in essence, introduces a single short{-}cut for two hops per pair
of overlapped hubs in the travel path. A typical travel path for this scheme
is indicated by the label $`I'$ in Fig.~3. As we will see in the following
section,  the introduction of this scheme, leads to an increase in the travel
speed and a reduction in the message delivery time. However, the actual
reduction in travel time for a given value of $k$, crucially depends on the
extent of overlap, which in turn depends on the minimum distance, $d_{min}$,
between hub nodes, as well as on the hub density. e.g. for the travel path
$`I'$ of Fig. 3, only one pair of hubs overlap so that the reduction in travel
time is just one unit.\\

The minimum distance between any pair of hubs, $d_{min}$, decides 
the range of overlap between the influence regions of the hubs
in our network. For any choice of $k$ at a fixed hub density, a
hub-distribution with $d_{min}=1$ on the lattice guarantees that the separation
between any pairs of hubs is equal to or greater than $1$. It is easy to see
that the areas of influence of a pair of hubs with $d_{min}=1$ have the
maximum overlap. See Fig.~1. On the other hand, $d_{min}=2k+1$ for a given $k$
results in no overlapping hubs on the lattice. Scheme~I can not be implemented
for this case. Distributions of hubs with other intermediate values of
$d_{min}$ have overlaps that lie between the two extremes.\\

\noindent {\bf Scheme~II}: In the second scheme to speed up the
communication process we connect individual hubs with a few (in this paper,
typically $2$ or $3$) other hubs selected at random. These connections, 
where nodes preferentially get connected  to nodes having a similar
degree of connectivity in the network, are called assortative connections
\cite{MayAlun,Newman2}. Under this scheme, when a hub becomes the current
message holder, it first tries to send the message through one of its
assortative linkages to another hub which, among all acqaintances of the hub,
happens to be the nearest to the target. If the current message holder hub
cannot utilise its assortative linkages because of unsuitable locations of the
end{-}point hubs, the message is sent to the constituent node nearest to the
target. A typical travel path between source $`S'$ and target $`T'$ is
indicated by the label $`II'$ in Fig.~3. Here the Manhattan distance $D_{st}$
between source and target is $142$ where the Manhattan distance is defined as
$D_{st}=|is-it|+|js-jt|$ and ($is$, $js$) and ($it$, $jt$) are the
co{-}ordinates for the source and the target, respectively. However the travel
path labelled $`II'$ needs just $50$ steps to travel between the source and
target. In comparison, the path labelled $`O'$ needs $95$ steps and that
labelled $`I'$ needs $94$ steps.\\

The simulations are carried out as follows. Two nodes are selected as the
source and the target at random from a lattice of  a given size. The
distance between them, denoted by  $D_{st}$, is chosen to be the {\it Manhattan
distance}. The number of steps required in delivery of the message from the
source to the target are counted for 50 realisations of hubs for a given hub
density. Then, the two nodes, $i.e.$, the source and the target nodes, are
replaced by two other nodes selected at random from the lattice, keeping
$D_{st}$ unchanged. Again the message transmission steps are counted for the
same number of hub realisations. This is repeated for 1000 pairs of source and
target nodes for a particular hub density. It should be noted that the order
of averaging makes no difference. The value of $d_{min}$ and $k$ used
throughout the paper are 1 and 3, respectively, unless otherwise specified.

\subsection{Cross-over behaviour}
The average travel time between a source and target a fixed distance apart is 
a good measure of the efficiency of the network for message transmission. 
This clearly depends on the density of hubs in the network, as well as on the 
way in which these hubs are connected. We study the behaviour of average travel
times as a function of hub density for a fixed Manhattan distance $D_{st}$
between source and target. Our simulations were carried out for a lattice of
$500 \times 500$ nodes $D_{st}=712$ for the original networks as well as the
networks modified by Schemes~I and II. Fig.~4 shows the dependence of the
average travel times, $t_{avg}$, as a function of hub density for the original
networks (diamonds) for $d_{min}=1$. The average travel times decrease
exponentially as the hub density increases. The data can be fitted well by the
exponential function \cite{FN1}, $f_1(x)=Q_1\exp[-A_1x^{\alpha_1}]$
where $\alpha_1=0.4482$, $A_1=0.0142$ and $Q_1=730$.
This can be rewritten in the form $f_1(x)=Q_1\exp[-(\frac{x}{x_0})^{\alpha}]$
where $x_0$ is $(\frac{1}{A_1})^{\frac{1}{\alpha_1}}$ and has the approximate
value 13259. Expanding this we get $f_1(x)=Q_1(1-X+\frac{X^2}{2!}-\cdots)$
where $X=(\frac{x}{x_0})^{\alpha_1}$. Retaining terms to the lowest order, we
see that the dependence of average travel times on hub density is given by 
$t_{avg}\approx Q_1(1-a \rho_{hub}^{\alpha_1})$, an instance of fat fractal
like behaviour. We plot the average travel times as a function of hub density
on the same plot for the Scheme~I and Scheme~II networks. The behaviour of
$t_{avg}$ in the case of Scheme~I networks (plus signs) is slightly different
from that of the original network case, and the exponential function acquires
a mild power{-}law correction. The data for Scheme I networks is fitted well
by a function $f_2(x)=Q_2\exp[-A_2x^{\alpha_2}]x^{-\delta}$ where
$\alpha_2=0.46$, $A_2=0.0145$, $Q_2=735$ and $\delta=0.00005$. Scheme~II
networks show  distinctly different behaviour. The Scheme~II network data
(boxes for two assortative connections and crosses for three assortative
connections) can be fitted by a function, $g(x)=Sx^{-\beta}$ where $\beta=0.2$
and $S$ is a positive constant and thus show power-law behaviour. We note that
the same power law fits both sets of Scheme II data.\\
 
Fig.~5 shows the same data as  Fig.~4 on a log-log plot. Here the  drastic
difference  seen for the behaviour of average travel times in the case of
networks with Scheme~II operational compared to the other two networks can be
clearly seen. The log-log plot of $t_{avg}$ against the hub density is
a straight-line with slope $\beta =-0.2$. Thus $t_{avg} \approx 
\rho_{hub}^{-\beta}$ shows power-law behaviour. It is again clear 
that the same power-law  is seen for the two Scheme II cases. Thus the addition
of a very small number of assortative connections per hub has induced a
cross-over to power-law behaviour from the fat fractal behaviour seen in the
non-assortative cases. The rate of fall off of average travel times with
increase in hub density is much faster for the assortative case even in the
case of low hub densities. Thus the addition of assortative connections can
increase the communication efficiency of networks without increasing the
number of hubs in the network especially at low hub densities.\\

We note that the cross-over from fat-fractal behaviour to power-law behaviour
is insensitive to perturbations of the regular lattice geometry. 
We verified that upto $10 \%$ variation in the $x$ and $y$ location in
the location of the hub-nodes makes no difference to the cross-over behaviour,
although the numerical values of the exponents change. Similar perturbations
to the location of all the nodes in the lattice also make no difference to the
existence of the cross-over. Changing the shape of the influence areas from
squares to circular regions introduces a variation in the degree of each hub,
and also a perturbation to the over-all degree distribution. We note that the
cross-over is robust to this change, as well. Fig. 5 (c) show the behaviour of
the hub density for the lattice network with a perturbed node distribution with
a circular influence area for the hubs, with all other parameters as in Fig.
5(a) and Fig. 5(b). It is clear that the cross-over is completely stable to
perturbation. The travel times can be fitted by the functions 
 $F(x)=635\exp(-0.0104 x^{0.417})$
for the perturbed  lattice with no assortative connections, and by the function
$G(x)=735x^{-0.165}$ for the perturbed case with assortative connections. We
note that the exponent $\beta$ now takes the value $0.165$ from the value $0.2$
seen earlier, and the power-law  $\alpha_1$ changes to $0.417$ from the value
$0.4482$ but the constant  $A_1$ barely changes from $A_1=0.0142$ to  
$A_1=0.0104$ a change in the third place after the decimal. Thus the
fat-fractal behaviour of the type $C(1- x^\alpha_1)$ remains unaltered due to
the perturbation as does the power-law behaviour for the network with
assortative connections. Thus, the cross-over is robust to perturbation.
\subsection{Distribution of travel times}
The probability distributions of travel times  on the network for the three
cases above can be seen in Figs 6 and 7. It is clear that  as hub densities 
increase, the peak shifts towards short travel times for the original network.
When the message was transmitted using Scheme~I, the distribution changes very
slightly with virtually no effect at low hub densities. However, when Scheme~II
was operational, the distribution showed a marked change (See Fig.~7). 
The distributions of Fig.~6 (the original lattice and Scheme~I) are sharply
peaked about a mean travel time and are symmetric about the mean whereas those
of Fig.~7 show a much wider spread and are skewed. There is also a difference
between the two distributions as shown in Fig.~7. At low hub densities the
distribution is bimodal, whereas the bimodality is smoothed out at higher hub
density. The right peak is reminiscent of the distribution for a lattice
without any hub{-}to{-}hub connections at  low hub density. Overall, the new
distribution  indicates that even sparse  hub{-}to{-}hub connections are
quite capable of inducing  short paths at all hub densities. 
The distribution also shifts to lower values of travel times demonstrating 
the success of the short-cutting assortative strategy (Scheme~II).
\section{Communication networks vs. spread of infections}
\label{Disease}
The message transfer problem discussed above involves the consistent
directed transfer of a message towards a target. Each temporary message holder
transfers the message, with probability one, in the direction which takes the
message towards the target. Other types of processes of information spread
such as the spread of computer viruses, infectious disease, rumours, 
popular fashions etc. have distinctly different mechanisms of spread. 
These processes are not directed processes, and incorporate stochastic 
elements in the mechanism of spread. Hence communication networks which may be
very efficient for information dissemination of the first type, may not be at
all, or less suitable, for the second kind. In this section we examine the
spread of an infection in a population of susceptible individuals for our
network. The population contains both infected and susceptible individuals.
The individuals constitute the nodes of the network, and social
interactions among them constitute the edges by which infection can be
transferred from one node to another. A similar model is also applicable to the
spread of computer viruses. We consider a single point entry for infection in
our study.\\

Many recent studies  have focussed, on this kind of spread of diseases on
networks  \cite{Watts-newman,Moore,Kuperman,Ziff,Newman1,Vespignani1,
Vespignani2}. A susceptible individual can  get infected (with some
probability)  only when he directly or indirectly encounters infectious
individuals in the population. The structure of the contact network has
important implications for three things. The first is the rate at which an
infection can spread across the network, the second is the transmission
threshold, $i.e.$ the smallest probability of infection, with which the
infection can spread to a significant fraction of the nodes of the network,
and the third is the  choice of an effective immunization strategy.\\
 
Recent studies of  disease spreading viruses or computer virus spreads on
networks of the SF type show that the possibility of these viruses
being persistent in the population, and of their being resurrected causing
repeated epidemics, is almost independent of any transmission threshold
\cite{Vespignani1}. Thus, in the case of the SF networks, all that is needed
for the infection to spread  throughout the population, is the occurrence of
a single-time point entry of the infection into the population. Since it has
been argued that in the case of sexually transmitted diseases, particularly
that of the HIV/AIDS infection, the underlying contact networks have scale-free
character \cite{Liljeros}, this result also has implications for the spread of
such diseases. However, the existence of a vanishing threshold in the case of
human disease does not fit well with the conventional epidemiologist's
view-point. The infinite variance of the node-connectivity in the SF network,
has been identified  as the causative agent for the vanishing threshold seen
in such networks \cite{MayAlun,AlunMay,Victor}. Thus, though traditional
epidemiology acknowledges the importance of heterogeneity in the rapid
spatial spread of diseases like HIV/AIDS, the extreme heterogeneity seen
in SF networks may not make them  good candidates for modelling
other kinds of infection in human or animal populations, where  social
contacts generally do not conform fully with the SF characteristic
\cite{Amaral,Holme}. Hence the study of disease spread on other types of
networks such as ours, which have finite variance, is important.\\

As mentioned above, network topologies also determine the choice of 
effective immunization strategies. Recent studies on immunization strategies
for efficient and successful controlling of epidemics on  SF networks have
unanimously advocated the immunization of the most connected nodes
\cite{MayAlun, Zoltan, Vespignani3}. The choice of immunization strategy would
have to be quite different in the case of networks where the node-connectivity
distribution is not of the SF type. Even in the case of SF networks,
immunization strategies which concentrate on immunizing the most connected
nodes do not take into account the finite probability that a few infected
nodes, which  may not be  the most connected nodes on the network, might have
long range connections to new regions of the susceptible population, and hence,
could  transmit the disease to regions quite distant from their place of
origin where a fresh epidemic can ensue. Our network, which incorporates
geographic distance and local clustering, is useful for such a study.\\

We studied information spread via an SIR type of  infection process on this
network. The process  has the following features: (i) all nodes are equally
susceptible, (ii) infection  always starts from a single site, (iii) an
infectious node can infect any of the nodes it is connected to with
probability $p$, (iv) this transmission probability is the same $i.e.$ it has
the value $p$, irrespective of whether the infected node is a hub or an
ordinary node, (v) none of the nodes get infected twice. Also once a node
becomes infected, it remains infected  until it infects any one of its
neighbours. Here we assume that within a typical infectious period an
infected node certainly infects, at least, one susceptible node. In this
study we restrict ourselves to a single episode of epidemic break.\\

We study the manner in which the infection spreads on a $100 \times 100$ node
lattice.  Fig. 8 shows the manner in which the infection wave-front from a
single infected node travels on a lattice with no hubs at $t=5$, $t=10$ and
$t=15$ (top row, labelled A), on a lattice with hubs (middle row labelled B)
at the same time steps , and a lattice with hubs with two assortative
connections (bottom row labelled C). The rapid spread of infection in the
bottom row can be very clearly seen. All the results in this section study
infection spread for a $100 \times 100$ lattice with $d_{min}=1$ and $k=3$.
Other parameters are given in the figure captions. Each run is taken for
$400$ time-steps.

\subsection{\bf Threshold behaviour and cross-over}
The threshold value of the transmission probability, $p_{th}$, is a very
crucial factor in the spread of disease. This quantity may depend on the
structure and topology of the network. In the case of the spread of computer
viruses on scale-free networks, the threshold value turns out to be zero,
leading to very rapid spread of virus across such networks. In the case of
immunological diseases, the threshold probability is finite. In this section
we examine the behaviour of this quantity for our networks as a function of
the hub density.\\

The threshold transmission probability, $p_{th}$, is defined in the following
way. For a fixed run of $t$ time steps, the threshold value  is defined to be 
the smallest value of the transmission probability for which at least 50$\%$ of the total number of susceptible sites are infected by half the run.
Fig.~9 shows the behaviour of $p_{th}$ against the number of hubs.
The diamonds show the observed behaviour for the original network and the
plus signs the behaviour for the Scheme~II network  when there are two
assortative (hub-to-hub) connections per hub. The  data for the original
network could  be fitted using a function,
$F(x)=D\exp[-Mx^{\zeta}](x+1)^{-\xi_1}+C_{1}$. The behaviour for the Scheme~II
networks (plus signs) was somewhat different and had to be fitted using
another function,
$G(x)=D\exp[-Mx^{\zeta}](x+1)^{-\xi_2}+C_2$, where $D=0.25$, $M=0.0054$,
$\zeta=0.975$, $\xi_1=0.065$, $\xi_2=-0.13$, $C_{1}=0.046$ and $C_{2}=0.036$.
Thus, there is a change in the scaling behaviour of the threshold probability
between the original network and the scheme~II case. However this cross-over
is a gentle change from one power-law to another, unlike the drastic change
seen for the behaviour of the average travel times for the same case. The
change is more pronounced at  low hub density as in the other case. The reason
as to why the cross-over is drastic in the case of the message transfer process
but is more gentle in this case of the infection spread, may lie in the fact
that in the case of disease spread, the variance appears to be  the crucial
quantity which determines the threshold behaviour. We plot the variance as a
function of hub-density in Fig. 10 for the original network (diamonds) as well
as the network with assortative connections (plus signs). It is clear that
there is very little difference between the behaviour of the variance in the
two cases. This appears to be the reason for the gentle cross-over.  
\subsection{\bf Immunization strategies} 
Studies of the SF networks have emphasised the importance of high connectivity
nodes, and immunisation strategies which immunise nodes of high connectivity
are the most successful. However, nodes of moderate connectivity can open up
new regions to the spread of infection if they are connected to nodes which
are far away.  Real{-}world epidemic events have always been effected by the
long{-}distance movements of causative agents into susceptible regions
\cite{Mary}. Some recent studies \cite{Ahmed, Anderson} provide evidence for
this. While studies of the SW networks take cognizance of the fact 
that that long range connections are important in the spread of information,
it is never clear what fraction of connections in a small world are really
long range, since the network is always stochastically generated using
re-wiring probabilities. Local clusters also play a crucial role in the
spread of infection.\\

We try to isolate the contribution of long range connections and that of the
local clustering property in information and disease spread on the network
with two assortative connections per hub by the use of different immunisation
strategies. Fig. 10 plots the number of new infections as a function of time
for this network. The plus signs indicate the the number of new infections for
the unimmunised network. The first immunisation strategy immunises the bonds
which connect hubs separated by a distance greater than or equal to $100$
lattice units (this is the  Manhattan distance between the two hubs). For this
lattice the average number of bonds of this type is about $17 \% $ of the total
number of bonds. However, the hub itself is not immunised and can infect the
local cluster. It is clear that the immunisation of the long range bonds causes
the number of new infections to decrease (see plot with diamonds), and also
cause the infection to spread more slowly. However, this effect is not
pronounced. The second immunisation strategy innoculates both the hubs which
are connected by such long-range bonds so that no infection travel from the
hub to any of the nodes connected to it. The number of new infections for this
case is plotted with boxes in the same figure. It is clear that the rate of
spread of infection slows down, the number of new infections peaks at a much
lower value, and the distribution develops a long tail. The last plot, viz.
the plot with crosses shows the number of new infections as a function of time
if the same number of hubs is immunized as in the last strategy, but if the
hubs are randomly chosen (i.e. the hubs chosen do not necessarily have long
range bonds). It is interesting to note that there is very little difference
between the two distributions. Thus, the local cluster which connects to other
local clusters (not necessarily distant ones) appears to play a more crucial
role in the spread of infection than the existence of long range bonds. This
is unlike the behaviour seen earlier in the case of small world networks.
Thus immunisation strategies which target arbitrary local clusters are as
successful as those which target local clusters with additional long range
bonds. We observed this behaviour at both low and high hub densities.\\
\section{Conclusions}
To summarise, in this paper we have studied information spread on a
two-dimensional communication network with nodes of two types, ordinary nodes
which are connected to their nearest neighbours and hubs which are connected
to all nodes within a certain range of influence. The degree distribution for
this lattice is bimodal in nature, and has finite variance. The average travel
time for directed message transfer between source and target on this lattice
shows fat fractal behaviour as a function of the hub density, however the
introduction of a small number of assortative connections between the hubs
(Scheme II) induces a cross-over to power-law behaviour for this average
travel time. In the case of Scheme I networks, where a short-cut was
introduced between end to end hubs  for consecutively overlapped hubs, a much
milder cross-over was seen. We also study the spread of infection on this
network by the SIR process. The threshold level for the infection probability 
is finite for the networks with and without assortative connections, due to
the fact that both networks have finite variance. However, the threshold level
as a function of hub density shows cross-over behaviour when assortative
connections are introduced when compared with the original network. However,
this cross-over is gentle in comparison to that observed for the average 
travel times for the directed message transfer for the same Scheme II case.
Thus, while network topology modifies the way in which information spreads on
a network, the effect appears to be stronger for directed processes than for
undirected processes. We also study the spread of infection and immunisation
strategies for this network, and conclude that local clustering plays as
important a role as the existence of  assortative connections in the rate of
spread of infection. Thus assortative connections play a more crucial role in
message transfer processes than in the spread of infection.\\

Our results can be of practical utility in a variety of contexts. In the case
of directed message transfer, at low values of hub density, the average travel
time between source and target can be reduced very rapidly by the
introduction of very few assortative connections per hub. This is a very
efficient way of reducing travel time without the introduction of new hubs.
Long range connections between hubs cut the travel time drastically in these
cases. On the other hand, the existence of a local clusters which can connect
to other local clusters (not necessarily distant ones) seems to play an
important role in the spread of infection. Thus immunization strategies which
target local clusters appear to be called for. It is thus important to note
that different elements of the network topology appear to be important for 
different types of information spread processes. We hope to explore this
direction further in future work.
\section{Acknowlegement}
We thanks CSIR, India for partial support for this work.

\newpage
\begin{figure}
\includegraphics[width=14cm]{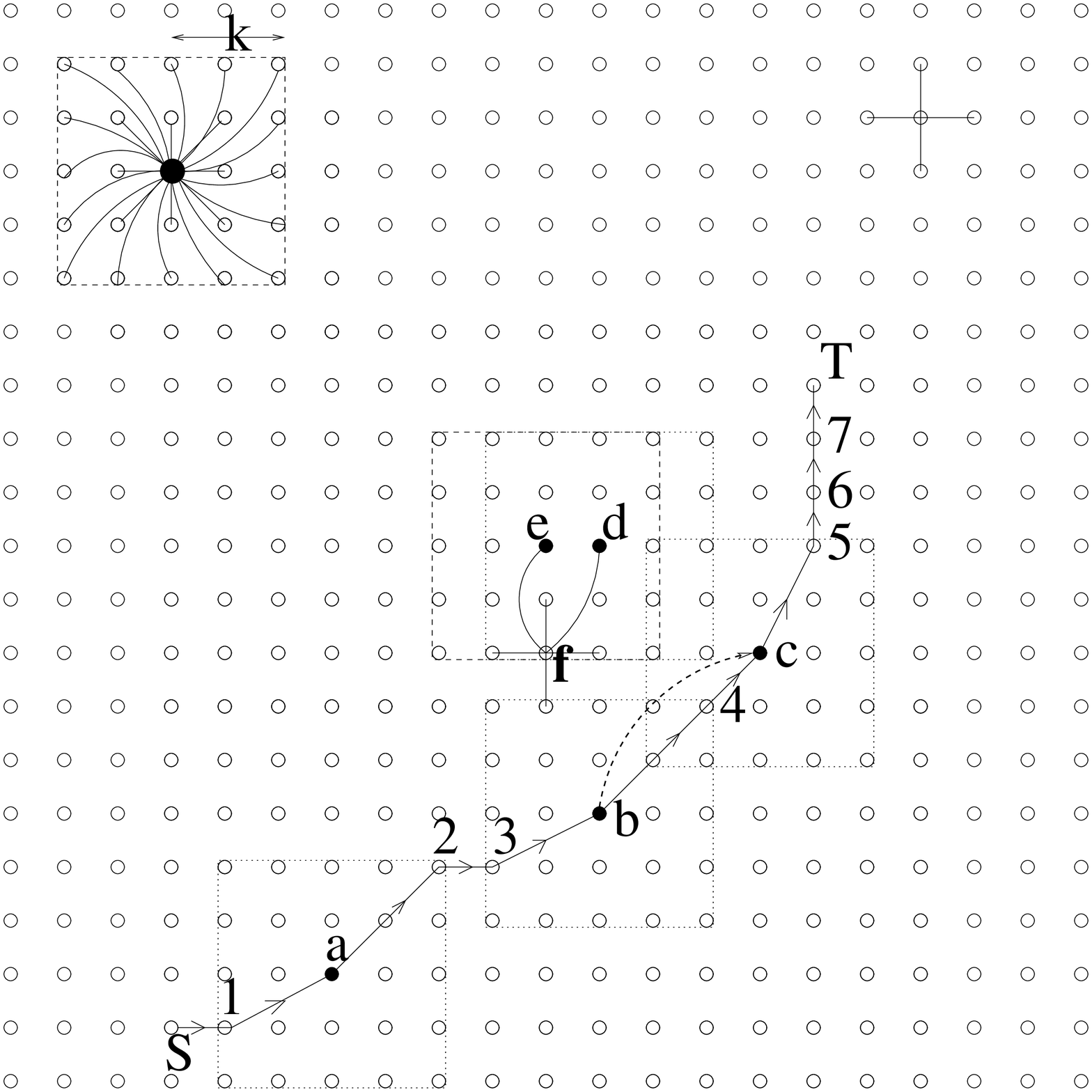}
\caption{\label{fig:epsart} A 2-dimensional lattice of ($21\times21$) nodes.
The top left corner shows a hub (a filled circle) which is directly connected
to its all constituent nodes within the square shown by the dashed lines.
The side of the square is approximately equal to $2k$, shown for $k=2$.
The square shown is called the influence area of  a hub. A regular node that
is neither a hub nor a constituent node has four edges connected to its
nearest neighbours (shown in the top right corner). A typical path  between
a source node, S, and a target node, T, is shown with the labelled sites.
Notice that it  passes through three hubs, namely, a, b, and c (filled
circles). After Scheme~I the distance between b and c is covered in one step.
The short{-}cut is shown by the dashed arrow from b to c.}
\end{figure}
\newpage
\begin{figure}
\includegraphics[width=11truecm]{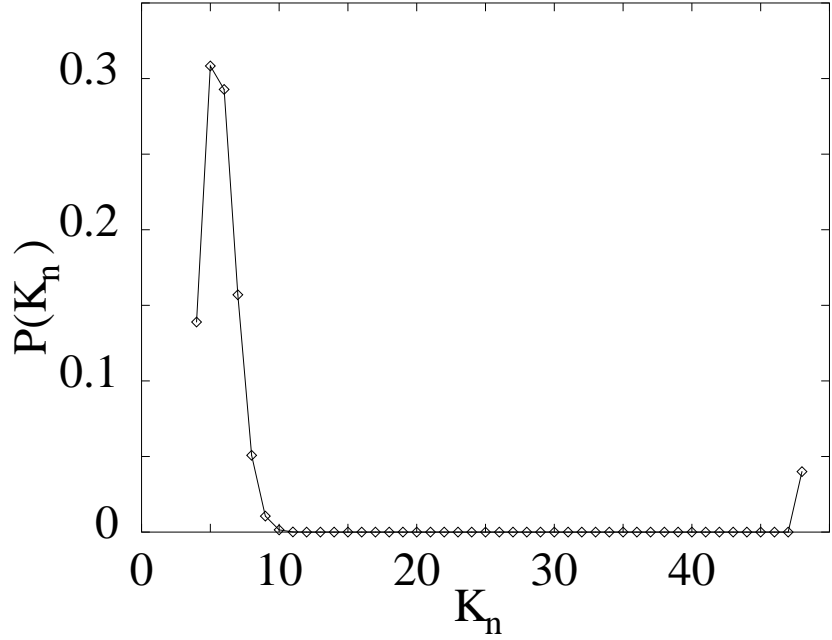}
\caption{\label{fig:epsart} A typical connectivity distribution for the
proposed communication network. $K_{n}$ gives the degree of connectivity
of a node with $K_{n}$ other nodes in the lattice. The lattice size is
($100\times100$), $k=3$ and the hub density is $4.0\%$.}
\end{figure}
\newpage
\begin{figure}
\includegraphics[width=11truecm]{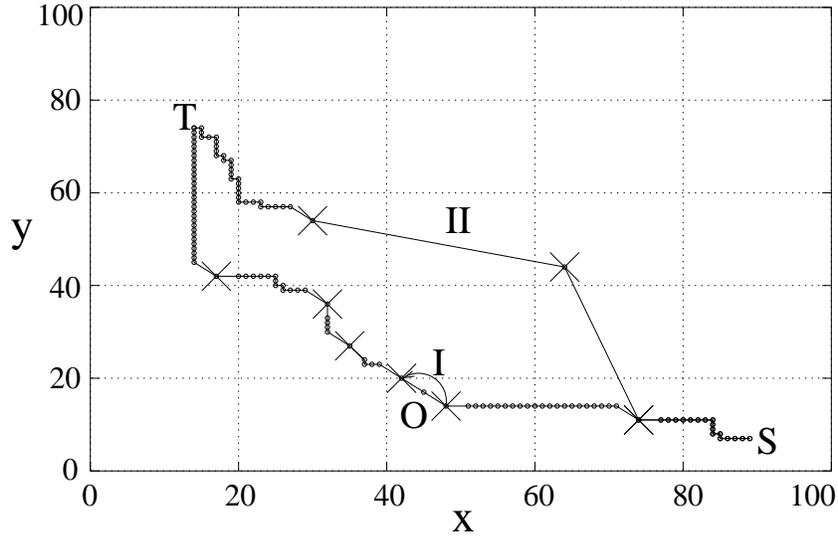}
\caption{\label{fig:epsart} Typical travel paths between source $S$ and
target $T$. The label `{\bf O}' indicates a path on the original lattice,
`{\bf I}' a path on the scheme I lattice and `{\bf II}' indicates a path
when the lattice was modified by the scheme II. The lattice distance
between for the source and target node is 142 LU. The  lattice size is
($100\times100$), $k=3$ and the hub density is $0.5\%$.}
\end{figure}
\newpage
\begin{figure}
\includegraphics[width=11truecm]{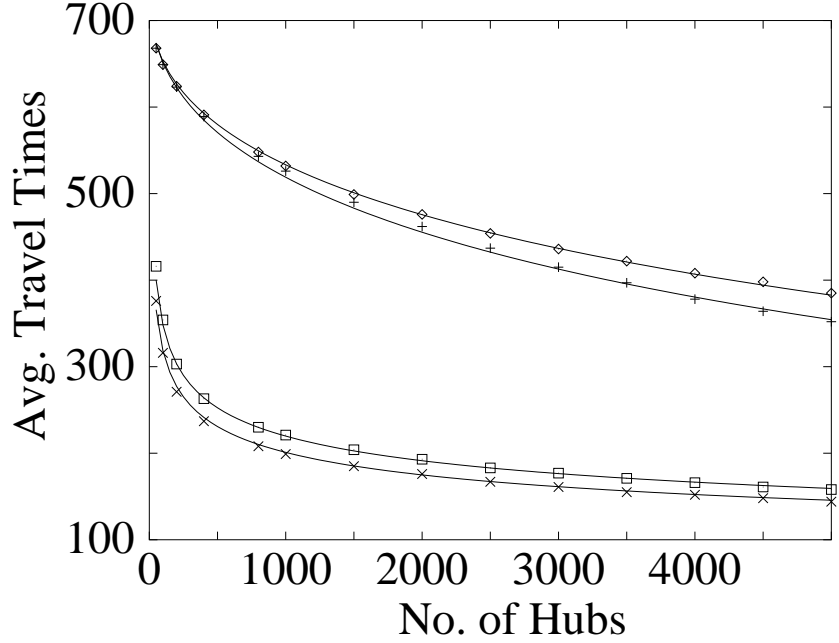}
\caption{\label{fig:epsart} A plot of average travel times, $t_{avg}$,
vs number of hubs for the original network (diamonds), the Scheme~I
network (plus signs) and the Scheme~II network (boxes) for two extra
assortative connections and (crosses) three extra assortative connections
per hub. Here, $d_{min}=1$. The best-fit line for the original network
(diamonds) was given by the function $f_1(x)=Q_1\exp[-A_1x^{\alpha_1}]$
where $\alpha_1=0.4482$, $A_1=0.0142$ and $Q_1=730$. The behaviour of
$t_{avg}$ for the Scheme~I network is slightly different and the
exponential function needs a mild power{-}law correction given by
$f_2(x)=Q_2\exp[-A_2x^{\alpha_2}]x^{-\delta}$ where $\alpha_2=0.46$,
$A_2=0.0145$, $Q_2=735$ and $\delta=0.00005$. For Scheme~II networks the
best{-}fit lines were given by the function, $g(x)=Sx^{-\beta}$ where
$\beta=0.2$ and $S=875$ (boxes), 800 (crosses).}
\end{figure} 
\newpage
\begin{figure}
\includegraphics[width=11truecm]{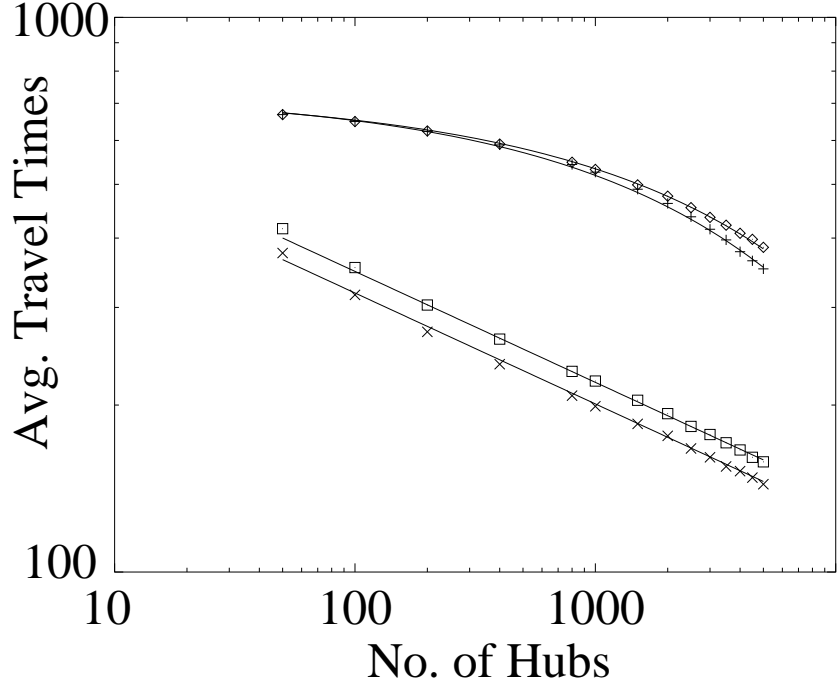}
\caption{\label{fig:epsart} The same data as in Fig.~4 is plotted  on the 
log-log scale. These plots clearly show cross-over in  scaling behaviour
from the fat fractal type seen for the original as well as the Scheme~I
networks to the power{-}law behaviour  for the Scheme~II network. Note that
the slope of the two parallel lines (the Scheme~II network) is $-0.2$.}
\end{figure}

\newpage
\begin{figure}
\includegraphics[width=11truecm]{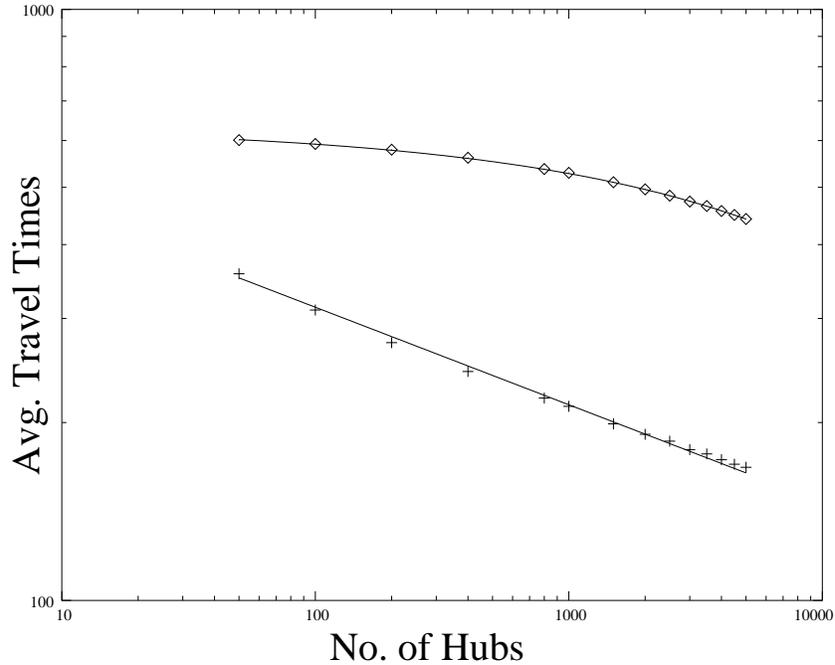}	
\caption{\label{fig:epsart} The parameter values are the same as in Fig. 5
except for the following. First, the influence area for this figure is
circular with a radius of $k$. Second, every node of the lattice network
has been displaced by $\pm 0.1$ from its earlier position as was in Fig. 5.
The functions for the fitted lines are : $F(x)=635\exp(-0.0104 x^{0.417})$
for $diamonds$, and $G(x)=735x^{-0.165}$ for $pluses$.}
\end{figure}

\newpage
\begin{figure}
\includegraphics[width=11truecm]{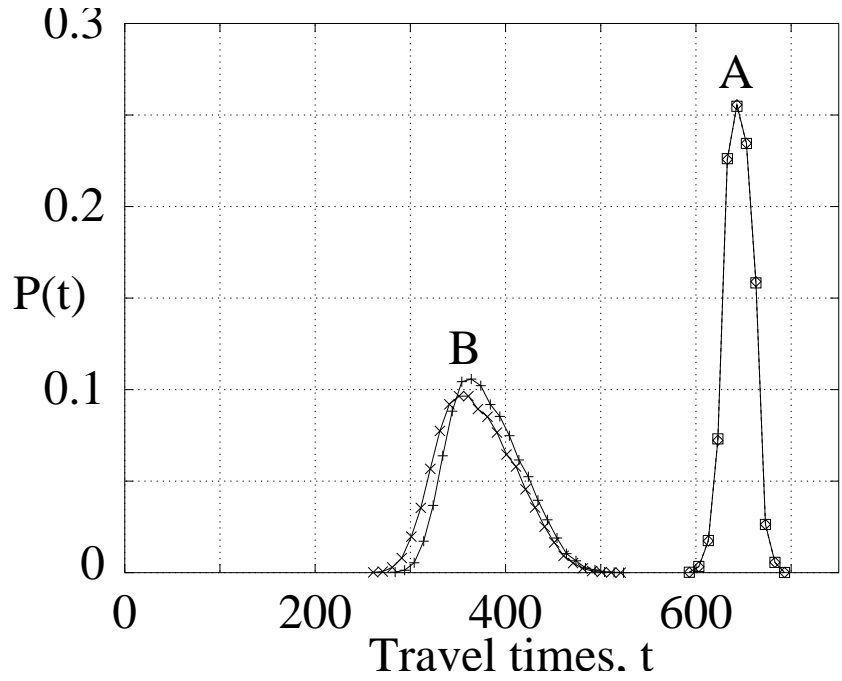}
\caption{\label{fig:epsart} The distributions of paths in terms of travel times
for $D_{st}$ equal to 712 on a lattice of (500 $\times$ 500). The distributions
shown are for $d_{min}=1$. The curve indicated by `I' is for a total of 100
hubs and the curve by `II' for a total of 5000 hubs. Plots with diamonds
and plus signs use   the original network data while the boxes and crosses
correspond to  Scheme~I data.}
\end{figure}
\newpage
\begin{figure}
\includegraphics[width=11truecm]{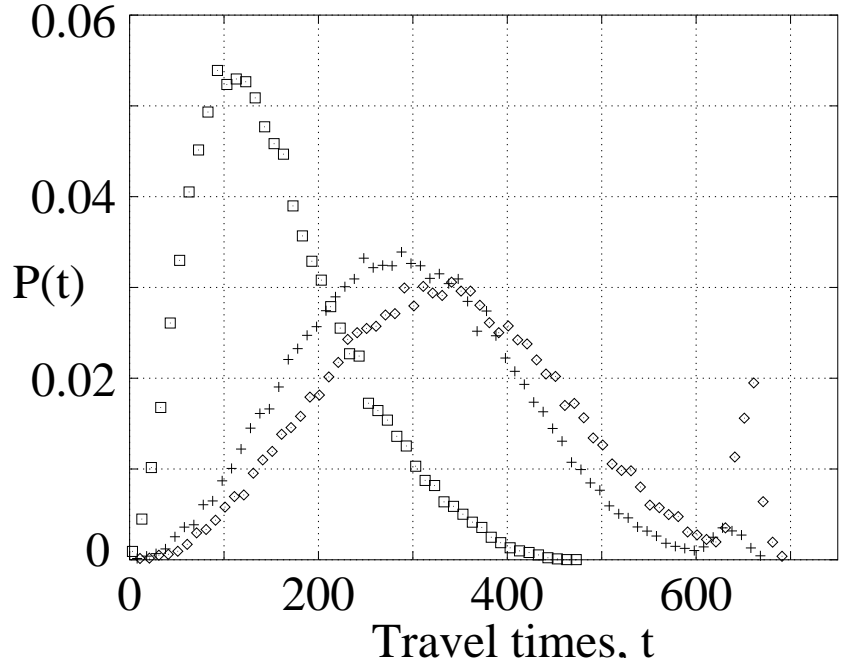}
\caption{\label{fig:epsart}
The distributions of paths in terms of travel times for the same value of
$D_{st}$ when Scheme~II is operational. Diamonds for a total of 100 hubs,
plus signs for a total of 200 hubs and boxes for a total of 5000 hubs.
The lattice size, $D_{st}$ and $d_{min}$ are the same as in Fig.~6. 
The distributions show bimodality (two peaks) at lower hub densities
(diamonds and plus signs), and this bimodality disappears at higher
hub densities (boxes).}
\end{figure}

\newpage
\begin{figure}
\includegraphics[width=11truecm]{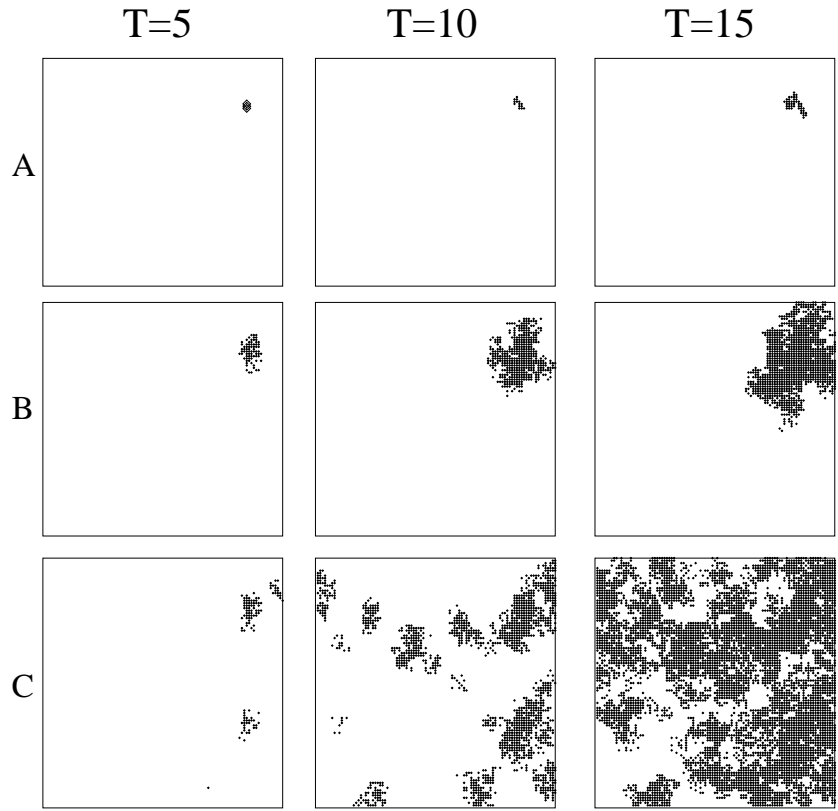}
\caption{\label{fig:epsart} 
The spread of infection on a lattice with no hubs (top row, labelled A), on
a lattice with hubs (middle row, labelled B), and a lattice with hubs with
two assortative connections (bottom row, labelled C). Snapshots of the
infection spread are taken at $t=5$, $t=10$ and $t=15$ (column labels) for
each case.}
\end{figure}

\newpage
\begin{figure}
\includegraphics[width=11truecm]{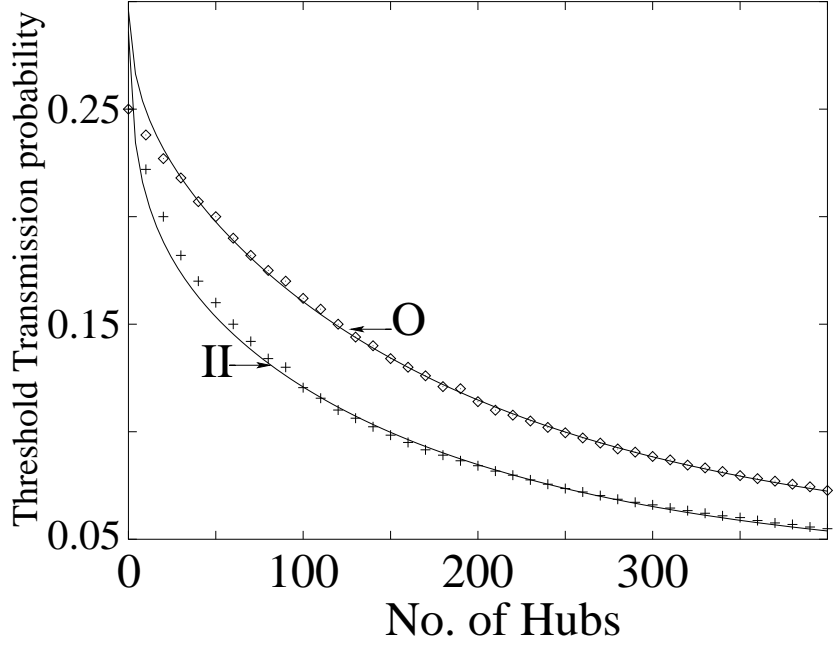}
\caption{\label{fig:epsart} The behaviour of the threshold values of
transmission probability, $p_{th}$, as the number of hubs increases.
The best{-}fit line for plot~I (for the original lattice)
was drawn using the function, $G(x)=D\exp[-Mx^{\zeta}](x+1)^{-\xi_1}+C_{1}$.
The behaviour for the lattices with two assortative  connections per hub,
shown by plot~II,  had to be fitted using another function,
$H(x)=D\exp[-Mx^{\zeta}](x+1)^{-\xi_2}+C_2$, where $D=0.25$, $M=0.0054$,
$\zeta=0.975$, $\xi_1=0.065$, $\xi_2=-0.13$, $C_{1}=0.046$ and $C_{2}=0.036$.}
\end{figure}
\newpage
\begin{figure}
\includegraphics[width=11truecm]{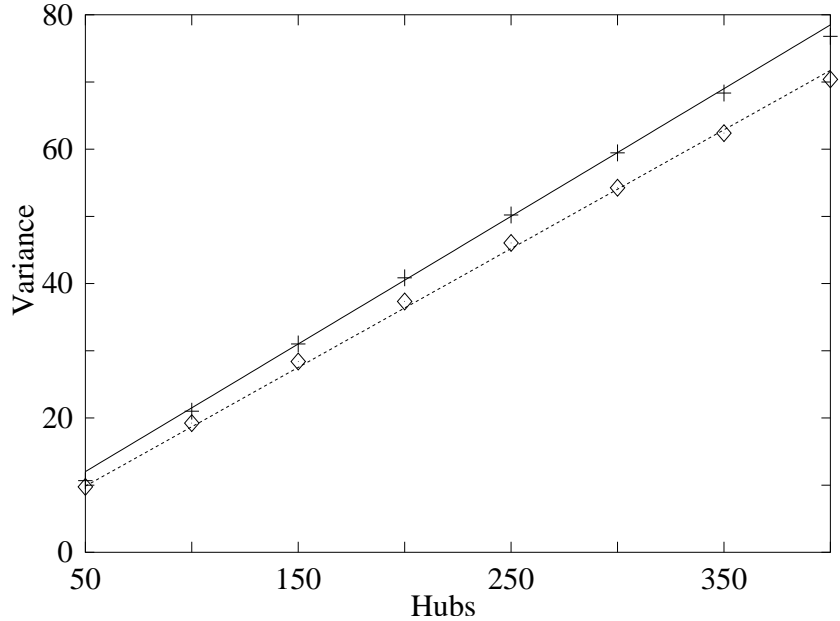}
\caption{\label{fig:epsart} The plot shows the variance in connectivity as
a function of hub density for two cases - the original (diamonds) and
scheme II networks (pluses). The lattice size is ($100\times 100$), $k=3$
and two assortative connections per hub for the Strategy II case.
}
\end{figure}

\newpage
\begin{figure}
\includegraphics[width=11truecm]{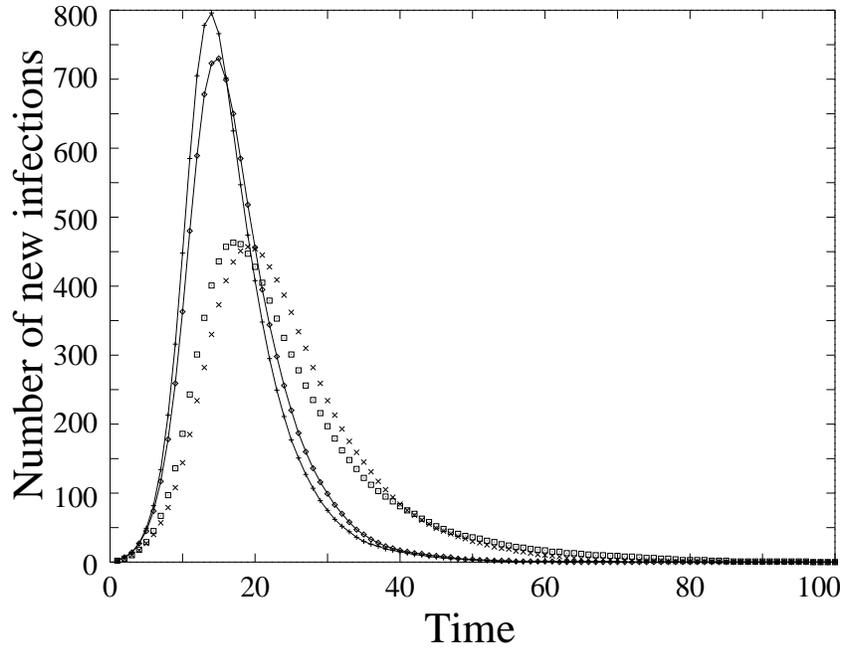}
\caption{\label{fig:epsart} The effect of immunisation  on the number of new
infections when  assortative bonds of length equal to or greater than $100$
LU are immunised (diamonds), when hubs with assortative bonds of length
greater than $100$ LU are immunised (boxes, $117$ hubs are immunised on an
average in this case) and when $117$ randomly chosen hubs are immunised
(crosses). The plot also shows the number of new infections as a function
of time (pluses) when the network has $400$ hubs with two assortative
connections per hub (the non-immunised case).
}
\end{figure}

\begin{thebibliography}{99}
\bibitem{FN}
It has been established that despite the regular geometry, traffic on
the network reproduces the characteristics of realistic internet
traffic\cite{Sole}
\bibitem{FN1}
Other values of $d_{min}$ lead to similar behaviour. The observed behaviour
can be fitted by $f_1(x)=Q_1\exp[-A_1x^{\alpha_1+(d_{min}-1)\epsilon}]$
where $\epsilon=0.005$ and other values are as in the text.
\end{thebibliography}
\end{document}